# A General Machine Learning-based Approach for Inverse Design of One-dimensional Photonic Crystals Toward Targeted Visible Light Reflection Spectrum

Tao Zhan [a,b,#], Quan-Shan Liu [a,b,#], Lu Qiu [a,b], Yuan-Jie Sun [c], Tao Wen [a,b], and Rui Zhang [a,b,*]

[a] South China Advanced Institute for Soft Matter Science and Technology, School of Molecular Science and Engineering, South China University of Technology, Guangzhou 510640, China
[b] Guangdong Provincial Key Laboratory of Functional and Intelligent Hybrid Materials and Devices, South China University of Technology, Guangzhou 510640, China
[c] School of Material Science and Engineering, South China University of Technology, Guangzhou 510640, China
[#] These authors contributed equally to this work
[*] Corresponding author
Email address: rzhang1216@scut.edu.cn (R. Z.)

**Abstract:** Data-driven methods have increasingly been applied to the development of optical systems as inexpensive and effective inverse design approaches. Optical properties (e.g., band-gap properties) of photonic crystals (PCs) are closely associated with characteristics of their light reflection spectra. Finding optimal PC constructions (within a pre-specified parameter space) that generate reflection spectra closest to a targeted spectrum is thus an interesting and meaningful inverse design problem, although relevant studies are still limited. Here we report a generally effective machine learning-based inverse design approach for one-dimensional photonic crystals (1DPCs), focusing on visible light spectra which are of high practical relevance. For a given class of 1DPC system, a deep neural network (DNN) in a unified structure is first trained over data from sizeable forward calculations (from layer thicknesses to spectrum). An iterative optimization scheme is then developed based on a coherent integration of DNN backward predictions (from spectrum to layer thicknesses), forward calculations, and Monte Carlo moves. We employ this new approach to four representative 1DPC systems including periodic structures with two-, three-, and four-layer repeating units and a heterostructure. The approach successfully converges to solutions of optimal 1DPC constructions for various targeted spectra regardless of their exact achievability. As two demonstrating examples, inverse designs toward a specially constructed "rectangle-shaped" green-light or red-light reflection spectrum are presented and discussed in detail. Remarkably, the results show that the approach can efficiently find out optimal layer thicknesses even when they are far outside the range covered by the original training data of DNN.

## 1. Introduction

Photonic crystals (PCs) possess periodic optical nanostructures which can be thought of as optical analogues of semiconductors[1–3]. Because of numerous extraordinary optical properties of photonic crystals, they have a wide range of applications in diverse fields, from traditional usage as components in optoelectronics and optical communications, to currently extended realms such as the environmental and energy technologies[4], biosensors [5], third-generation photovoltaic cells [6], integrated sensing platform [7], anti-fouling [8], and smart strain-color sensing [9]. Investigation of fundamental physics, as well as materials design and optimization of photonic crystals, thus represents an important scientific topic in optical science and technology.

Depending on their geometrical symmetry, PCs can be generally divided into three classes, namely one-dimensional (1D), two-dimensional (2D), and three-dimensional (3D) structures. The simplest 1D case is of enduring research interest due to its relatively easy construction and convenient fabrication as well as the ability to still possess a rich set of intriguing optical properties [10]. 1D PCs (1DPCs) consist of alternating layers of material with different dielectric constants. In this sense, 1DPCs include all stacked layer-by-layer structures which exhibit certain regularity along one direction. Since the optical properties of 1DPCs are closely associated with their reflection spectra, it is of significant value to explore inverse design methods that can guide the construction of 1DPCs toward different desired reflection spectra. In particular, reflection spectra at visible-light frequencies have always been of high practical relevance. For example, quite a few dielectric materials are nearly lossless in the visible light region, therefore the optical loss problem can be neglected. Many practical applications of 1DPCs are also based on the visible-light reflection spectra, such as sensors[7,11], environmental and energy technologies [4], fluorescence enhancement [12], and full-color displays[13].

Traditionally, the development of optical materials and systems relies on physics-inspired methods and human intuition which provide the guidelines to the material and structure design. The emerging data-driven methods, such as machine learning, open new avenues for optics research. Machine learning can solve specific problems that cannot easily be tackled by conventional experimental or forward computational methods. Notable recent examples include a machine learning-based approach to design a PC gas sensor for identification and estimation of three greenhouse gases [14], a high-accuracy convolutional neural network for band structure prediction with orders-of-magnitude speedup compared to conventional theory-driven solvers [15], a tandem neural network structure for overcoming the issue of nonuniqueness in the inverse scattering of electromagnetic waves [16], an artificial neural network-based method to approximate light scattering by multilayer nanoparticles [17], a deep learning-based model for accelerated all-dielectric metasurface design [18], a physics-driven neural network for global optimization of dielectric metasurfaces [19], and a neural network-based method to design optical structures with target topological properties of the Zak phase [20]. Despite these signs of progress, methodologies of machine learning-based inverse design of 1DPCs are still in their infancy, especially at a level to be generally effective.

In this paper, we propose a general inverse design approach for 1DPC systems that combines forward transfer matrix calculation and backward prediction and optimization based on deep neural network (DNN) [21]and Monte Carlo (MC) methods. The core functionality of the approach is to inversely find optimal 1DPC constructions within a pre-specified parameter space that yield minimal mean squared error toward a targeted visible light reflection spectrum. Four representative 1DPC systems, including periodic two-layer, three-layer, four-layer structures, and one heterostructure [2,22], are comprehensively studied. The results show consistent high accuracy across all systems and all types of targeted spectra, thus strongly supporting the general applicability of the approach.

The remaining parts of this paper are organized as follows. Section 2 introduces models and methods. We first give a brief introduction of 1DPC architecture and describe details of the four classes of 1DPC systems selected to test the performance of the inverse design approach. We then present key formulas of the transfer matrix method and the unified structure of DNN as well as its training process based on data from transfer matrix calculations. Next, we demonstrate technical details of the iterative algorithm, which integrates forward transfer matrix calculations, backward DNN predictions, and adjustments of the reference spectrum through MC moves to search optimal 1DPC constructions toward a targeted spectrum. Section 3 presents numerical results at each stage. We first focus on visible light reflection spectra derived from forward transfer matrix calculations, which naturally belong to "exactly achievable" spectra since 1DPC constructions corresponding to the original transfer matrix calculations are the exact solutions. We show the trained DNNs can accurately predict backward the 1DPC layer thicknesses within 1nm precision on average over the test set. We then move to the optimization study toward targeted spectra based on the iterative inverse design algorithm. For both exactly achievable spectra and artificially designed spectra, we show the algorithm can always successfully finds the optimal solutions of 1DPC constructions. The paper concludes with a summary and a discussion of future directions in section 4.

## 2. Models and methods

Figure 1 summarizes key model and method ingredients revolving around the general machine learning-based 1DPC inverse design approach proposed in this work. Clarification of details for each ingredient is given below.

### 2.1. 1DPC structures, four representative systems, and the transfer matrix method

1DPCs possess layer-by-layer structures where each layer is composed of a specific type of dielectric material. In this broad sense, a wide variety of 1DPC systems can be conceived, differentiated by the types of dielectric materials, the regularity of repeating units, the order of layer arrangements, and so on. Our goal is to establish a general and concise machine learning-based inverse design framework to be applicable to a rich range of 1DPC systems. With this purpose in mind, four representative 1DPC model systems (figure 2) are chosen as the testbed. There is no any calculation beforehand to bias our choices of systems. We emphasize that even if an inverse design strategy is consistently successful for all four systems (as we will show later, our approach indeed achieves such a level of performance), we will not claim it as a rigorous proof of the absolute applicability of the strategy to all classes of 1DPC systems. On the other hand, it certainly strongly supports the promise of the strategy to make valuable inverse predictions (from spectrum to 1DPC construction) for 1DPC systems not tied to a special category [23], thus greatly facilitating related 1DPC research and development due to its high generality. To the best of our knowledge, no machine learning-based study that pursues high generality of 1DPC inverse design has been reported in literature.

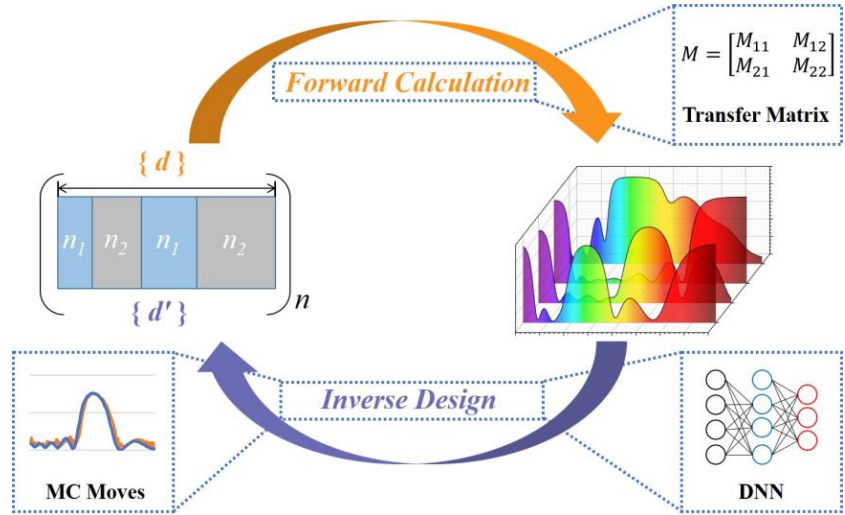

Fig. 1. Schematic of the computational methods implemented in this work. The forward calculation refers to the calculation of 1DPC visible light reflection spectrum based on the transfer matrix method. For each 1DPC system considered in this study, a sizeable number of forward calculation results are taken as machine learning data to train a system-dependent DNN in a unified structure. An iterative approach that combines DNN backward predictions (from reflection spectrum to 1DPC layer thicknesses), forward calculations, and MC moves forms a general, effective inverse design approach for finding optimal 1DPC constructions toward a targeted reflection spectrum.

We now turn to detailed information on the four 1DPC systems. To make our model calculations concrete at some common dielectrics, four types of materials are considered: TiO2, polymethyl methacrylate, Ta2O5, and polystyrene. In the remaining discussion, we use letters A, B, C, and D to represent them respectively. The corresponding refractive indexes are 2.4, 1.49, 1.97, and 1.59. In addition, we choose glass as the substrate, which has a refractive index of 1.52.

The first model (referred to as System 1 in figure 2) is a two-layer, AB-type 1DPC with six periods which has alternating A and B layers in one period. The term "two-layer" means the minimum period of 1DPC consists of two categories of materials with different refractive indices. The second model (referred to as System 2 in figure 2) is a three-layer, ABC-type 1DPC with four periods which has alternating A, B, and C layers in one period. The third model (referred to as System 3 in figure 2) is a four-layer, ABCD-type 1DPC with three periods which has alternating A, B, C, and D layers in one period. While the first three systems enrich the 1DPC diversity by qualitatively varying the basic architecture of the repeating unit (highlighted in black boxes in figure 2), they all belong to conventional 1DPCs since they are strictly periodic. To further increase the 1DPC diversity, thus making the generality of our final inverse design approach more convincing, we also consider a model (referred to as System 4 in figure 2) is a heterostructure. It is a Het-AB-CD-type system, which can be viewed as a stack of a two-layer, AB-type 1DPC with three periods and a two-layer, CD-type 1DPC with three periods. Finding optimal layer thicknesses of a 1DPC system toward a targeted reflection spectrum is the central inverse design problem studied in this paper. The numbers of independent thickness variables for the four systems are 2, 3, 4, and 4, respectively.

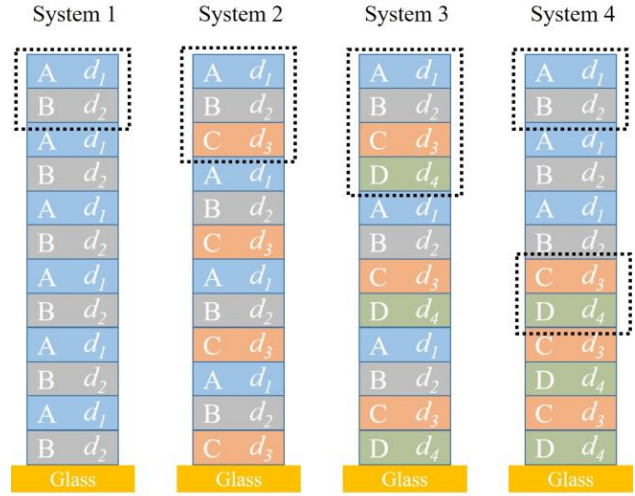

Fig. 2. Four classes of 1DPC systems: AB-category (System 1), ABC-category (System 2), ABCD-category (System 3), and Het-AB-CD category (System 4). The letters A, B, C, and D represent four types of materials with different refractive indexes. The total number of dielectric layers is fixed at 12 for all systems studied in this work. Highlighted in black square boxes are basic repeating units for each system.

In our approach, we refer to "forward calculation" as the calculation of reflection spectrum for a given 1DPC construction (thus all thicknesses are known). In all case studies discussed in this paper, forward calculations are conducted at normal incidence using the transfer matrix method. It should be noted that other methods to calculate reflection spectrum can also be chosen, so the generality of our inverse design approach is not restricted by any particular forward calculation algorithm.

The key ingredients of the transfer matrix method are briefly outlined below. For any given layer j, the transfer matrix $M_j$ can be expressed as:

$$M_j = \begin{bmatrix} \cos(kn_j d_j) & -\dfrac{i\sin(kn_j d_j)}{n_j} \\ -in_j \sin(kn_j d_j) & \cos(kn_j d_j) \end{bmatrix} \quad (1)$$

with the vacuum wavevector $k$, the refractive index of a material $n_j$, and the layer thickness $d_j$. Each layer can generate a transfer matrix, and the total transfer matrix is the product of all of them:

$$M = \prod_j M_j = \begin{bmatrix} M_{11} & M_{12} \\ M_{21} & M_{22} \end{bmatrix} \quad (2)$$

The reflection coefficient $r$ and the reflectivity $R$ can be calculated as follows:

$$r = \frac{M_{11}\eta_0 + M_{12}\eta_0\eta_s - M_{21} - M_{22}\eta_s}{M_{11}\eta_0 + M_{12}\eta_0\eta_s + M_{21} + M_{22}\eta_s} \quad (3)$$

$$R = r^* \cdot r \quad (4)$$

Here, $\eta_0 = n_0\sqrt{\varepsilon_0}/\sqrt{\mu_0}$ and $\eta_s = n_s\sqrt{\varepsilon_0}/\sqrt{\mu_0}$. $\varepsilon_0$ and $\mu_0$ are vacuum permittivity and vacuum permeability while $n_0$ and $n_s$ are the refractive indices of air and glass (two ambient media in this paper), respectively.

## 2.2. Deep neural network

A DNN that predicts backwards the 1DPC layer thicknesses based on input reflection spectrum is an important tool in our inverse design framework. Several considerations are relevant to our construction of the DNN structure. First, since we only pay attention to reflection spectra in the visible light window with wavelengths in the range from 380 nm to 780 nm, we standardize DNN input parameters as 81 discrete reflectance values at wavelengths starting from 380nm and ending at 780nm, with a 5nm regular increment in between. We note that the 5nm resolution is good enough to capture delicate spectrum features for design purposes. A much finer spectrum resolution (thus much more input parameters) is not recommended as it will strongly increase the computational cost while gaining little extra benefit. Second, the number of output parameters (1DPC thicknesses) is much smaller than 81. For the four systems, it is in the range between 2 and 4, which is typical for many 1DPC systems. The strong difference between the numbers of parameters in the input and output layers suggests that the number of neurons in each hidden layer can be designed in a decreasing trend (from input to output layer) to reduce computational cost while maintaining accuracy. Third, the number of hidden layers should properly reflect the complexity of 1DPC systems. This is a common consideration in deep learning research, since too many and too few hidden layers take the risk to overfit and underfit the training data respectively.

After some numerical experiments guided by the above considerations, we finally build a unified DNN structure as shown in figure 3. Specific implementations are realized by Keras machine learning library in TensorFlow [24]. Our unified DNN has an input layer (81 input nodes), an output layer (the number of output nodes is the same as the number of independent 1DPC thicknesses), and four hidden layers. The number of nodes (neurons) in each hidden layer is (from input to output layer) 256, 128, 64, and 16, respectively.

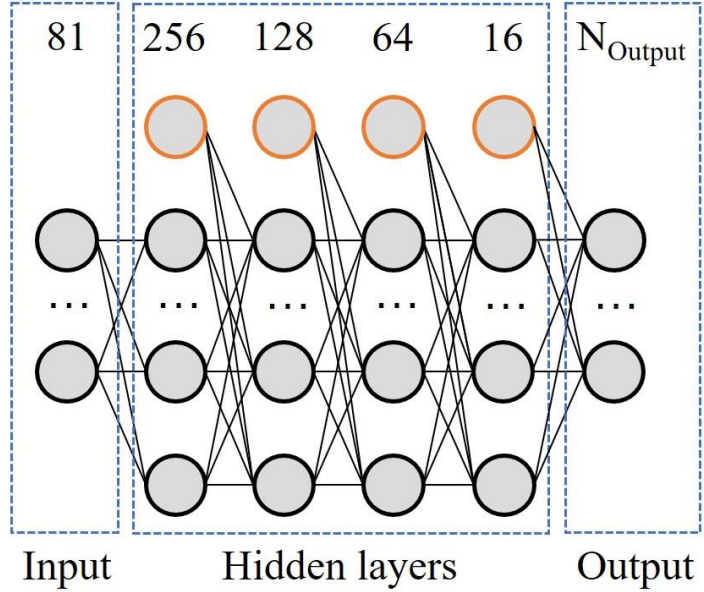

Fig. 3. Schematic of the unified fully-connected DNN structure. The input parameters are 81 reflectance values of a given visible light reflection spectrum. The outputs (backward predictions) are layer thicknesses for a specific 1DPC system ($N_{Output}$ = 2, 3, or 4). The DNN contains four hidden layers. The numbers of neurons in each hidden layer are labeled in the figure. The orange circles represent additional bias units (one in each hidden layer).

Below we describe detailed DNN training processes for the four 1DPC systems. The training, validation, and test data are all obtained from forward transfer matrix calculations. Regarding the thickness range of these data, we partially refer to the quarter-wave stack condition. Adopting this condition, the major lower and upper bounds (40nm and 115nm respectively) of the thickness ranges are determined by those of the visible light waveband. For System 1 which has only two independent thicknesses, we set the thickness range as 40nm-200nm. For all other three systems, the thickness range is set as 40nm-115nm. Regarding the amount of data, the total number of spectra generated for System 1 is 102 400, where the two materials A and B each randomly take 320 values in a uniform distribution within the thickness range. System 2 randomly takes 75 values within the thickness range for each type of material, and the total number of spectra generated is 421 875. System 3 and System 4 both randomly take 38 values within the thickness range for each type of material, and the total number of spectra generated is 2 085 136 for both systems.

The proper density of training data needs some comments. Our study indicates that a 2nm thickness resolution is stable to maintain accuracy, in a sense that higher resolution certainly gives a smaller average error of the DNN backward prediction, but the relationship is roughly linear. However, too sparse data, such as 5nm thickness resolution, generate a strong error. We conclude that 2nm thickness resolution or higher is a rule that one should follow. The resolution of our training data for each system is 0.5nm, 1nm, 2nm, and 2nm, respectively.

The data usage of the DNN training and test is distributed as follows. 80% of the data is used for training, the other 10% is used as the validation set, and the remaining 10% is used as the test set. The activation function of the input layer is tanh and others are rectified linear units (ReLU). The thickness mean-squared error (MSE) is adopted as the loss function. The training of the DNN model is based on mini-batch gradient descent, and the size of each batch is 128. Each DNN model has been trained for 2000 epochs. An epoch refers to one iteration of the entire training set.

## 2.3. Iterative optimization of 1DPC thicknesses toward targeted reflection spectrum

A high-quality DNN can make backward predictions close to the optimal solution by just a one-step calculation, but to further increase the accuracy of inverse design, additional algorithms need to be created. This need motivates the iterative optimization method introduced here. We emphasize that a trained DNN is an essential component of this algorithm, so it is an important precondition.

For the convenience of discussion, we first define several concepts. The input spectrum $S$ of DNN is called *original spectrum*. The DNN outputs are called *predicted thicknesses*. From the transfer matrix calculation, setting 1DPC thickness values as the predicted thicknesses will get another reflection spectrum $\hat{S}$, which is called *predicted spectrum*.

Equation (5) expresses the above relations in a concise fashion:

$$\hat{S} = M[D(S)] = f(S) \tag{5}$$

where $S = \{y_1, y_2, \ldots, y_n\}$ and $\hat{S} = \{\widehat{y_1}, \widehat{y_2}, \ldots, \widehat{y_n}\}$ represent the group of discrete reflectance values (in the range of [0, 1]) of the original spectrum and predicted spectrum respectively (in our standard format for visible light spectrum, $n$ is fixed at 81 as introduced in section 2.2), M represents the transfer matrix method, D represents the trained DNN, and f denotes the net function that converts S to Ŝ. In our study, the error between the original reflection spectrum and the predicted reflection spectrum is quantified by the reflectance MSE,

$$MSE = \frac{1}{n}\sum_{l=1}^{n}(\widehat{y_l} - y_l)^2 \tag{6}$$

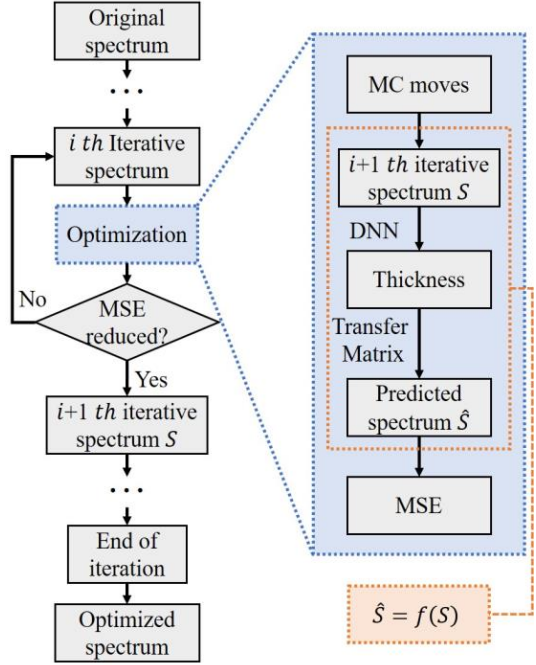

Fig. 4. The workflow of the optimization (inverse design) process for a targeted initial reflection spectrum. Details for each step are described in the main text.

Our iterative optimization scheme is summarized in figure 4. Here, *MC moves* refers to randomly selecting several reflection spectrum points and randomly changing their reflectance values within a certain range. This allows DNN to explore more spectrum features, thereby approaching more accurate backward predictions. Since the first prediction of DNN in the thickness training range is generally accurate, the underlying "landscape" is not too underfitting to disallow the possibility to find the optimal solution. Another advantage of this iterative optimization method is its universality. It is applicable to any 1DPC system, and the iteration convergence speed will not be significantly reduced as the 1DPC systems become more complex, because this method is iteratively optimized based on the reflection spectrum instead of 1DPC system parameters such as the thicknesses.

Detailed steps of the iterative optimization algorithm are as follows:

1. Input the targeted reflection spectrum to function $f$, obtain the initial iterative spectrum, initialize the iterative process;

2. Apply MC moves to the previous iterative spectrum to get a new iterative spectrum;

3. Calculate the predicted spectrum by equation (5) corresponding to the new iterative spectrum;

4. Calculate the reflectance MSE between the predicted spectrum and targeted spectrum by equation (6);

5. If the reflectance MSE decreases, accept the new iterative spectrum generated in Step 2, otherwise, reject the MC moves in Step 2;

6. Iterate Steps 2 to 5 until the reflectance MSE is converged to the minimal (optimal) value, output the optimal 1DPC thicknesses.

To ensure the reflectance MSE is not trapped at some possible local minima, the MC moves of Step 2 should change the reflection values of multiple spectrum points at a time. In our study, we always perform four independent iterative calculations following the workflow in figure 4 by making 2, 5, 10 or 20 MC moves in Step 2, and finally select the minimal reflectance MSE case among the four as the optimized inverse design result (corresponding reflection spectrum is called *optimized spectrum*). Even though in most situations, all four iterative calculations converge to the same result, we still recommend taking the "four independent calculations" as a general rule as it is robust enough to guarantee the successful finding of the optimal spectrum toward the targeted original spectrum. Regarding the reflectance variation range in each MC move, a general rule concluded from our studies is to randomly choose a number in a uniform distribution in the interval of [-0.3, 0.3] and add it up to the initial reflectance value. Because the values of reflectance are physically constrained in the range of [0, 1], if the added sum is larger than 1 (smaller than 0), we apply a natural rule to set the new reflectance value just as 1 (0).

## 3. Results and discussion

### 3.1. DNN training: generality and accuracy test

In figure 5, the thickness MSE is used as the loss function to measure the DNN training process, which is an average over many different samples in the validation set. The DNN training errors of the four 1DPC systems decrease with an increasing number of epochs and all have converged after 2000 epochs. The verification set error is very close to the training set error. The errors of the training set and the validation set are ultimately less than 1.5 nm2 for all cases. The numerical results of the loss function show no over-fitting phenomenon. System 2 has over-fitting in the early stage of training, but as the number of training epochs increases, the thickness MSE finally converges.

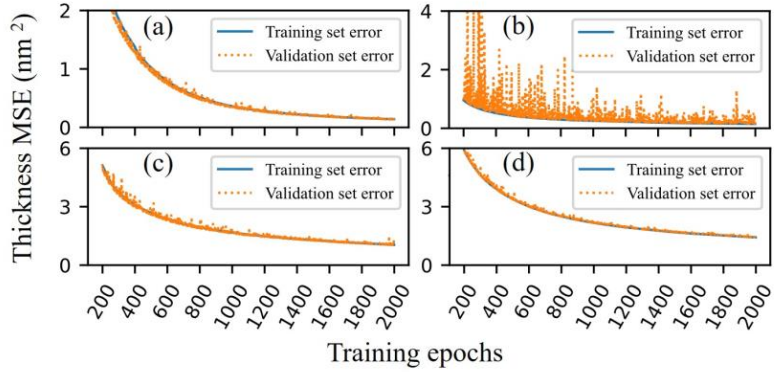

Fig. 5. The records of the DNN training process (thickness mean squared error as a function of the number of training epochs) for (a) System 1, (b) System 2, (c) System 3, and (d) System 4.

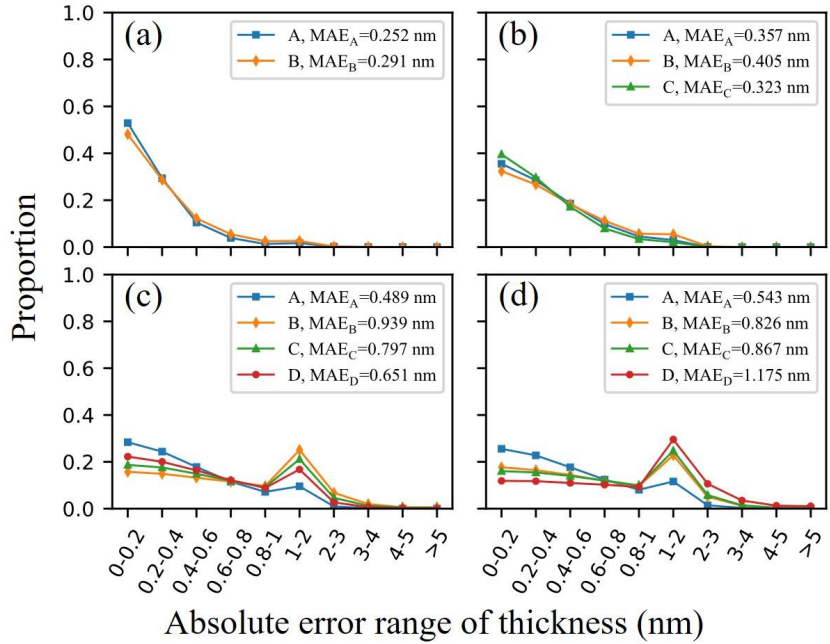

Fig. 6. The distribution of thickness absolute errors (divided into ten intervals as indicated by the x-axis label) of DNN backward predictions on the test set for (a) System 1, (b) System 2, (c) System 3, and (d) System 4. Different symbols represent results for different types of dielectric layers. The thickness MAEs for each type of layer are shown in legends.

The trained DNN is used to predict the thicknesses on the test set. As shown in figure 6, the predicted values of thicknesses are in most cases very close to the true values. Due to the uneven distribution of the x-axis scale, the curves present a protrusion at "1-2" in figure 6(c) and figure 6(d), but if the proportions of absolute errors are less than 1nm (hence the first 5 proportion values on each curve) are added up, the resulting total proportion corresponding to the "0-1" range always dominates. The legends of figure 6 summarize the thickness MAEs for all types of materials and systems. Recall the training data density (see details in section 2.2) follows the order: System 1 > System 2 > System 3 = System 4. As expected, higher training data density indeed yields smaller thickness MAEs. The overall thickness MAEs of DNN predictions over the test set for all four systems are below 1nm, indicating a good level of accuracy.

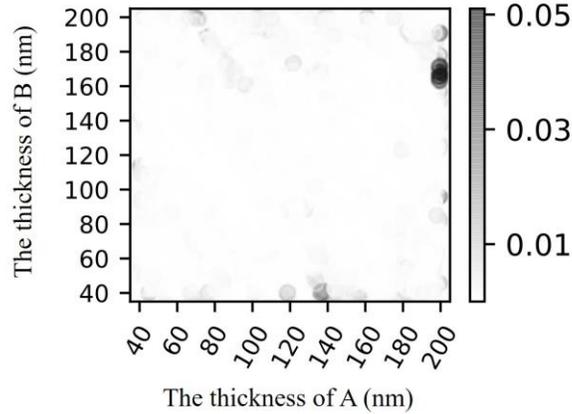

Fig. 7. The reflectance MSE (magnitude indicated by the grey-scale legend) between the original and predicted spectrum as a function of original thicknesses of A and B layers over the test set of System 1. For a given original spectrum, the corresponding predicted spectrum is obtained via two steps: first, the DNN model converts the original spectrum to predicted thicknesses of all layers; secondly, the transfer matrix method is used to calculate the predicted spectrum based on the predicted thicknesses.

The reflectance MSE between the original spectrum and predicted spectrum is another measure of the DNN accuracy over the test set. Figure 7 shows the reflectance MSE (magnitude indicated by the grey-scale legend) as a function of the original thicknesses of A and B layers over the test set of System 1. We choose to show the results for System 1 just because of the convenience of plotting. Similar features are also exhibited on the other three systems. Most reflectance MSE values are invisible in figure 7 since based on our statistics results 99.5% of them are less than 0.01. The maximal value among them is 0.051. The overall low values of reflectance MSE again support that the trained DNN has the ability to predict thicknesses with high accuracy.

Figure 7 also clearly presents the feature that relatively large reflectance MSE values mostly occur at the edges of the thickness range in the training set. For example, it is visibly obvious (darkest spot) that the maximal reflectance MSE occurs at a place where the thickness of material A is close to 200nm. This edge effect is perhaps not surprising since the data density at the edges is lower than the central region. Table 1 lists the information of the maximal reflectance MSE cases over the test set for all four systems. The edge effect again emerges on System 2, System 3, and System 4, as at least one original thickness value is close to 40nm or 115nm for all of them.

Table 1. The maximal reflectance MSE between the original and predicted spectrum over the test set (second column) and the corresponding original thicknesses of different types of layers (third column) for the four systems.

| System | Max MSE | Original thickness |
|---|---|---|
| 1 | 0.051 | (199.38, 168.22) |
| 2 | 0.070 | (114.34, 114.41, 113.39) |
| 3 | 0.157 | (40.50, 51.45, 44.77, 41.47) |
| 4 | 0.056 | (65.33, 48.33, 100.08, 113.40) |

*3.2. Optimization and inverse design toward targeted reflection spectrum*

To test the ability of the iterative optimization method introduced in section 2.3, we first use it to optimize the maximum reflectance MSE cases listed in table 1. Note that the original spectra corresponding to these cases are "exactly achievable" since all test data are generated by the forward transfer matrix method (see section 2.2), meaning the original thicknesses used to obtain these original spectra are naturally the optimal (and also exact) solutions of the inverse design. Maximal reflectance MSE indicates the largest deviation of a one-step DNN prediction over the test set. If for such "worst" situations the iterative optimization method can successfully find the optimal solutions, it is strong support of the applicability of the iterative algorithm to inversely design "exactly achievable" reflection spectrum.

The optimization results for the four such "worst" cases are shown in figure 8, and indeed in each case the predicted spectrum based on a one-step DNN prediction deviates a lot from the targeted original spectrum. On the other hand, the iterative optimization calculations successfully converge to the optimized spectra for all four cases, indicated by the essentially zero value of reflectance MSE (see figure 8 legends) and perfect collapse between the original and optimized spectrum curves.

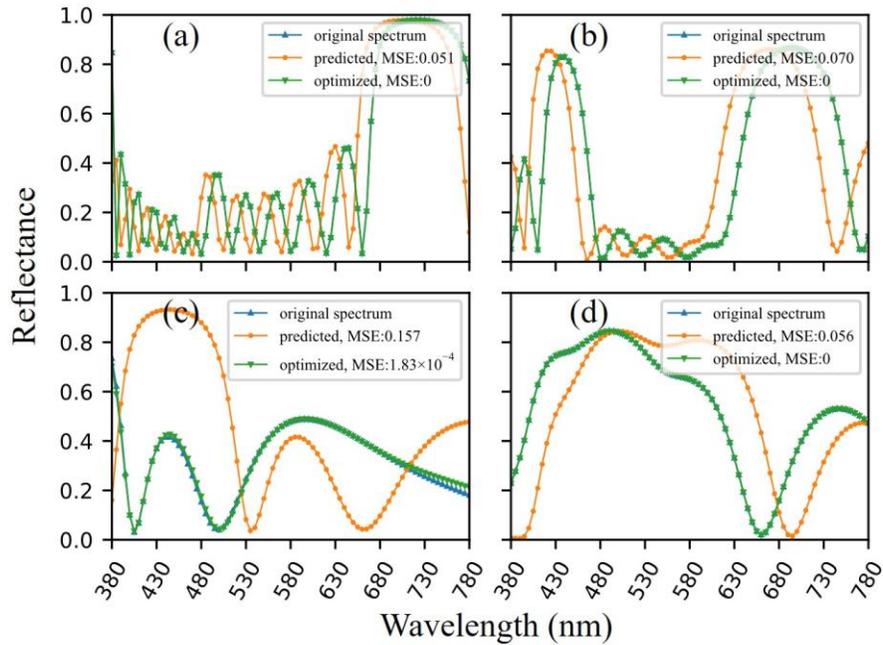

Fig. 8. The optimization results for the four maximal reflectance MSE cases are listed in table 1 based on the inverse design method: (a) System 1, (b) System 2, (c) System 3, and (d) System 4. The original spectrum (blue, triangle up) and the optimized spectrum (green, triangle down) strongly overlaps for all cases, consistent with zero or very small reflectance MSEs indicated in the legends.

The next interesting question to ask is whether the iterative optimization algorithm can also find optimal inverse design solutions for artificially constructed spectra. Here, "artificially constructed" means the targeted reflection spectrum is pre-specified without any prior forward calculations. Within the parameter space for a specific type of 1DPC system, the optimized parameters (what the inverse design pursues) satisfy the condition that the reflectance MSE between the spectrum they generate and the targeted spectrum attains the minimum achievable value, which is generally not zero for artificially constructed targeted spectra.

We have tested the ability of our iterative optimization algorithm for the inverse design of several different artificially constructed spectra and observed good performance consistently. Figure 9 presents an example of the inverse design for a targeted 1DPC that exclusively reflects green light. The wavelength range of green light is 500-570nm. We construct a "rectangle-shaped" spectrum as the target in which the reflectance values in the green light wavelength range are all set to 0.99, and in other ranges are all set to 0.01. The iterative optimization algorithm is applied to all four 1DPC systems, and the resulting optimized spectrum curves are plotted in figure 9. Figure 9 also shows three other types of spectrum curves, among them the meanings of original spectrum and predicted spectrum are clear based on the definitions given in section 2.3. Besides, the whole iterative optimization procedure is also repeated for a red light "rectangle-shaped" spectrum, with a wavelength range between 620 and 750nm. The result is illustrated in Figure 10. Below we explain the meaning of ergodic spectrum and its usefulness as a reference to check the quality of the optimized spectrum prediction.

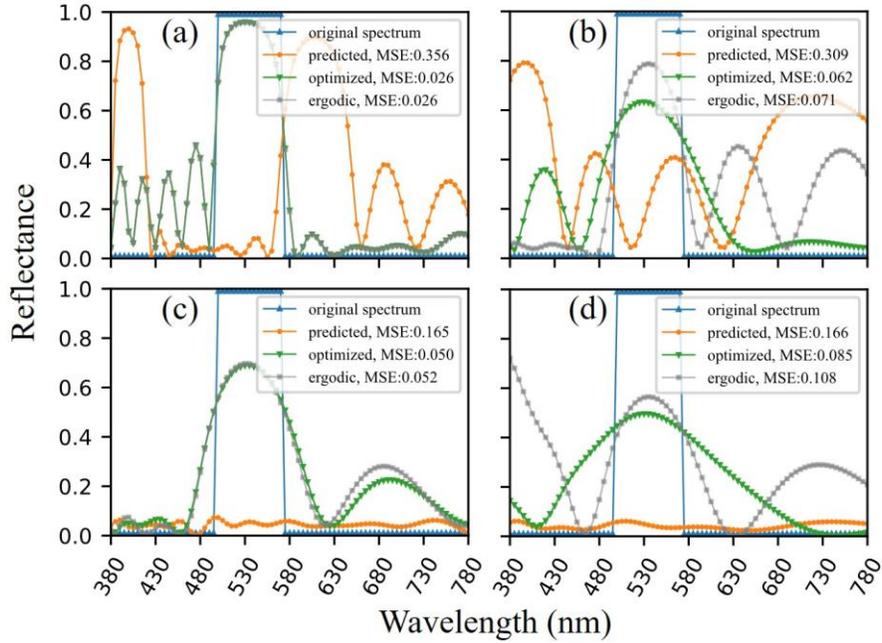

Fig. 9. The optimization results toward an identical targeted original spectrum (R=0.99 for wavelengths in a green light region (500-570nm) and R=0.01 otherwise) based on the inverse design method for (a) System 1, (b) System 2, (c) System 3, and (d) System 4. The original spectrum (blue, triangle up), predicted spectrum (orange, circle), optimized spectrum (green, triangle down), and ergodic spectrum (grey, square) are compared, and the corresponding reflectance MSEs are indicated in the legends.

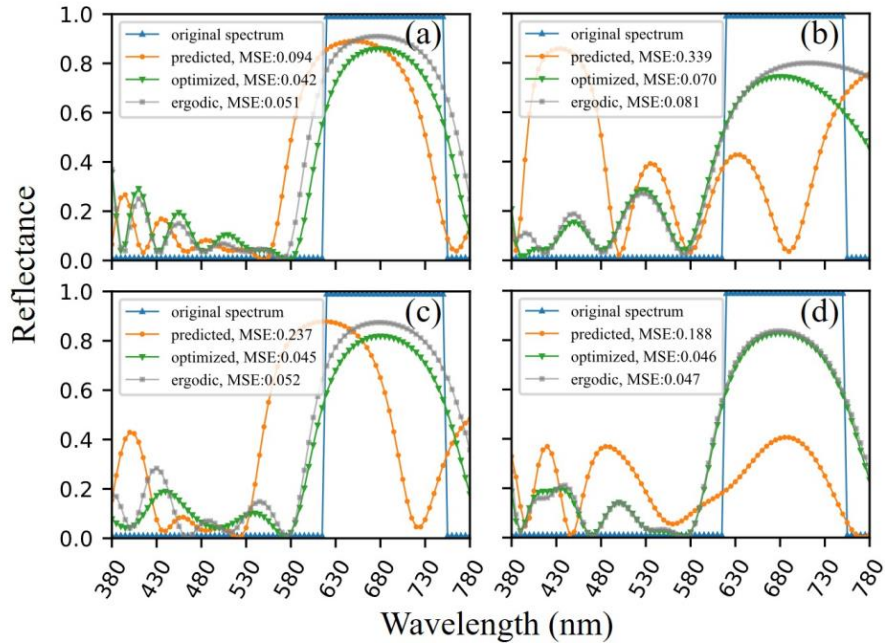

Fig. 10. The optimization results toward an identical targeted original spectrum (R=0.99 for wavelengths in a red light region (620-750nm) and R=0.01 otherwise) based on the inverse design method for (a) System 1, (b) System 2, (c) System 3, and (d) System 4. The original spectrum (blue, triangle up), predicted spectrum (orange, circle), optimized spectrum (green, triangle down), and ergodic spectrum (grey, square) are compared, and the corresponding reflectance MSEs are indicated in the legends.

One way to locate 1DPC parameters that generate a reflection spectrum close to the targeted spectrum is based on a brute-force searching method. In the case of the four 1DPC systems studied in this work, this method first chooses a thickness range, and then traverses different independent thickness values in this range and applies the forward transfer matrix method to calculate corresponding reflection spectra. Finally, all these spectra are compared with the targeted spectrum to determine the best values of thicknesses that yield the smallest reflectance MSE. The best spectrum obtained by this way is called ergodic spectrum.

Since the brute-force method exhaustively searches all possible thickness parameters in the pre-set thickness range, it ensures the ergodic spectrum is the optimal solution confined in that thickness range. In both figures 9 and 10, the ergodic spectrum curves for the four 1DPC systems are obtained through the brute-force method. Specifically, we set the thickness range for each system exactly the same as the range expanded by the data used in the DNN training. The density of thickness values in the brute-force search is also the same as that in the DNN training. The only difference is that the thickness values in the brute-force search are regularly spaced over the entire thickness range.

Since the iterative optimization method does not confine thickness values in any pre-set range (of course physically they must be non-negative), it holds the potential to find, outside the thickness range considered by the brute-force method, achievable spectra that give smaller reflectance MSE values than the ergodic spectrum. The results presented in figures 9 and 10, together with those in tables 2 and 3, clearly support this very valuable ability owned by the iterative optimization method. The reflectance MSE of the optimized spectrum is smaller than that of the ergodic spectrum for each system except System 1 in figure 9 and System 4 in figure 10, and indeed the thickness values corresponding to these optimized spectra are outside the pre-set range for the brute-force search (see tables 2 and 3). For System 1 in figure 9 and System 4 in figure 10, the iterative algorithm essentially converges at the ergodic spectrum. These results demonstrate the strong ability of the iterative scheme to effectively inversely design 1DPC constructions toward a targeted, artificially constructed reflectance spectrum.

**Table 2.** The layer thicknesses corresponding to the optimized spectrum (second column) and ergodic spectrum (third column) shown in figure 9 for the four systems.

| System | Thicknesses (optimized spectrum) | Thicknesses (ergodic spectrum) |
|---|---|---|
| 1 | (140.49, 134.62) | (140.5, 134.5) |
| 2 | (22.51, 46.97, 82.40) | (93, 43, 115) |
| 3 | (94.17, 12.90, 82.43, 71.01) | (98, 40, 52, 72) |
| 4 | (17.64, 155.39, 0, 182.64) | (40, 40, 82, 74) |

**Table 3.** The layer thicknesses corresponding to the optimized spectrum (second column) and ergodic spectrum (third column) shown in figure 10 for the four systems.

| System | Thicknesses (optimized spectrum) | Thicknesses (ergodic spectrum) |
|---|---|---|
| 1 | (114.37, 32.26) | (110, 40) |
| 2 | (14.63, 163.99, 41.48) | (91, 40, 40) |
| 3 | (92.52, 167.82, 17.01, 97.89) | (102, 40, 114, 88) |
| 4 | (112.27, 36.99, 98.02, 108.67) | (112, 40, 92, 114) |

Admittedly, "no worse than the ergodic spectrum" is not a rigorous proof that the optimized spectrum is the best answer in the complete thickness parameter space. Because such a space is infinite, there is hardly an exact way to prove an achievable spectrum has attained the minimal reflectance MSE over all parameter possibilities. We believe the fact that the iterative optimization method proposed in this work always successfully finds an achievable spectrum no worse than the ergodic spectrum is appealing enough to claim its effectiveness in the inverse design of 1DPC systems.

From the machine learning perspective, why the iterative algorithm can converge at a spectrum associated with thickness values far away from the original thickness range (System 4 in table 2 is a clear example) expanded by the DNN training data is a very intriguing question. It seems reasonable to worry about the accuracy of DNN guidance in the iterative optimization process when the thickness values are far outside the range of the original DNN training. After all, no data around these thickness values directly participate in the learning process of DNN. Our interpretation of this puzzle is as follows: the underlying physical relations between the thicknesses and the reflectance spectrum are universal (such as the transfer matrix formulations). Therefore even though the data used in DNN training only cover a limited thickness range, the DNN should have learned many features of the physical relations through these data, which are sufficient to ensure the DNN backward predictions at least follow correct physical trends regardless of the thickness values. By integrating the respective merits of DNN calculations (provide "spectrum-to-thicknesses" reverse predictions in a fast and at least qualitatively accurate fashion), MC moves (achieve an exploration of the spectrum space), and forward calculations based on the transfer matrix method (provide absolutely accurate "thicknesses-to-spectrum" results and accordingly the reflectance MSE values to measure the optimization process), our iterative optimization method stands out as a powerful inverse design tool. The combinatory and machine learning-based nature of the approach is of central importance governing its strong inverse design capability.

Finally, we briefly discuss two interesting system-dependent features. First, table 2 shows that the third thickness value of the optimized spectrum for System 4 is zero. Recall this value represents the thickness of the type-C layer in the heterostructure of System 4. Referring to figure 2, if all type-C layers are essentially absent (zero thickness), the resulting structure is actually a three-period AB-type 1DPC stacked on a single thick layer (thickness=3×182.64=547.92nm) composed of type-D material. The fact that the iterative algorithm can converge at such an extreme optimal solution is strongly supportive of its high robustness. Second, while not a focus of this study, we note that figures 9 and 10 also present the feature that different types of 1DPC systems have distinctive reproducing abilities toward an identical targeted spectrum. Cross comparing the optimized spectra of the four systems toward the artificially constructed "green-light reflection spectrum", System 1 obviously presents a much better performance as shown in figure 9. Thus System 1 is more suitable to realize the green light reflection effect among the four system choices. Likewise, for the "red-light reflection spectrum spectrum" case, System 1 is also the best option, though with little progress compared to Systems 3 and 4. The unified nature of our inverse design framework has the advantage of easily being applied to multiple distinctive types of 1DPC systems and selecting the overall optimal solution.

## 4. Conclusions

In conclusion, a general machine learning-based inverse design approach for 1DPCs is proposed in this work. To elaborate the working mechanisms and general applicability of the approach, four types of broadly representative 1DPC systems, including traditional two-layer, three-layer, and four-layer periodic structures and non-traditional heterostructures, are studied in detail.

The computational framework of the approach is built via two major steps. In the first step, a DNN in a unified structure is trained over a sizeable group of thicknesses-spectrum data from forward calculations (the transfer matrix method is used to obtain forward calculation data in this work). On the test set (reflection spectra therein are exactly achievable), direct DNN predictions are largely very accurate, but the performance is typically worse for the extreme cases at the edges of the data set.

The optimization method established in the second step takes the trained DNN as an indispensable component, and integrates the DNN backward predictions, MC moves and forward calculations into a powerful iterative inverse design workflow. The iterative algorithm successfully reduces the reflectance MSE to zero for all exactly achievable targeted reflectance spectra. To demonstrate its applicability to the inverse design of artificially constructed reflection spectra, we discuss two detailed examples where a spectrum with reflectance of 0.99 in either the green or red light wavelength range and 0.01 otherwise is constructed as the inverse design target. For all four 1DPC systems in both cases (eight case studies), the iterative algorithm successfully converges at an optimized spectrum with reflectance MSE equal to or smaller than that of the ergodic spectrum (best result confined in the thickness range of the DNN training data). Optimal thickness parameters are found far outside the DNN training data range in five out of eight case studies. In particular, an extreme optimal solution is found for System 4 in figure 9 in which the optimal thickness of the type-C layer is predicted to be zero. The capability to explore thickness parameter space far beyond the original range expanded by the DNN training data is an exciting merit of our iterative optimization method.

The inverse design approach proposed in this paper is valuable along several avenues in future 1DPC research. Due to its high generality, one can conveniently apply our approach to study specific 1DPC systems to his/her interest. For the purpose to inversely design new 1DPCs toward peculiar reflection spectrum patterns [25–29], which is of high practical interest, a promising route is to apply our unified iterative optimization algorithm to multiple candidate 1DPC systems and find out the overall optimal solution. By incorporating additional advanced machine learning techniques, such as the reinforcement learning method, our approach is hopeful to be generalized to a more flexible and functional inverse design tool in the future.

**Declaration of competing interest**

The authors declare that they have no known competing financial interests or personal relationships that could have appeared to influence the work reported in this paper.

**Funding**

This work was supported by National Natural Science Foundation of China (grant number 21973033) and the Fundamental Research Funds for the Central Universities (grant number 2018ZD13).